\documentclass[aps,preprint,showpacs,showkeys,nofootinbib]{revtex4}
\begin{document}

\preprint{Yukawa Institute Kyoto} \preprint{YITP-08-78}

\title{Simple unified derivation and solution of Coulomb, Eckart and \\
Rosen-Morse potentials in prepotential approach}

\author{Choon-Lin Ho
}

 \affiliation{Yukawa Institute for Theoretical Physics,
Kyoto University, Kyoto 606-8502, Japan\\and\\Department of
Physics, Tamkang University, Tamsui 251, Taiwan, Republic of
China\footnote{Permanent address}}

\date{Sep 30, 2008}

\begin{abstract}

The four exactly-solvable models related to non-sinusoidal
coordinates, namely, the Coulomb, Eckart, Rosen-Morse type I and
II models are normally being treated separately, despite the
similarity of the functional forms of the potentials, their
eigenvalues and eigenfunctions.  Based on an extension of the
prepotential approach to exactly and quasi-exactly solvable models
proposed previously, we show how these models can be derived and
solved in a simple and unified way.
\end{abstract}


\maketitle

\section{Introduction}

Exactly-solvable models are important in any branch of physics.
They allow a complete understanding of the dynamics of the
corresponding systems, and thus serve as paradigmatic examples for
the field of studies concerned.  However, exactly-solvable systems
are rather scanty, and hence any means to find them are always
welcome. It is thus very interesting that most exactly-solvable
one-dimensional quantum systems can be obtained in the framework
of supersymmetric quantum mechanics, based only on the requirement
of shape invariance \cite{Cooper}. Recently, the number of
physical systems which we can study analytically has been greatly
enlarged by the discovery of the so-called quasi-exactly solvable
models \cite{TU,Tur,GKO,Ush,PT,KMO}. These are spectral problems
for which it is possible to determine analytically a part of the
spectrum but not the whole spectrum.

In \cite{Ho} a unified approach to both the exactly and
quasi-exactly solvable systems is presented. This is a simple
constructive approach, based on the so-called prepotential
\cite{CS,Ho1}, which gives the potential as well as the
eigenfunctions and eigenvalues simultaneously. The novel feature
of the approach is the realization that both exact and quasi-exact
solvabilities can be solely classified by two integers, the
degrees of two polynomials which determine the change of variable
and the zero-th order prepotential.  Most of the well-known
exactly and quasi-exactly solvable models, and many new
quasi-exactly solvable ones, can be generated by appropriately
choosing the two polynomials. This approach can be easily extended
to the constructions of exactly and quasi-exactly solvable Dirac,
Pauli, and Fokker-Planck equations.

The exactly solvable models that are generated by the prepotential
approach in \cite{Ho} are related to the so-called sinusoidal
coordinates \cite{OS}. These are coordinates $z(x)$ whose
derivative-squared $z^{\prime 2} (x)$ (henceforth the prime
denotes derivative w.r.t. the variable $x$) is at most a quadratic
polynomial of $z$ (or equivalently, $z^{\prime\prime}$ is at most
linear in $z$). Such coordinates include quadratic polynomial,
exponential, trigonometric, and hyperbolic types. The
corresponding exactly solvable models are the six systems listed
in \cite{Cooper}: the shifted oscillator, three-dimensional
oscillator, the Morse, Scarf type I and II, and generalized
P\"oschl-Teller potentials. However, the other four models in
\cite{Cooper}, namely, the Coulomb, Eckart, Rosen-Morse type I and
II models cannot be generated by the approach as presented in
\cite{Ho}. These four models are based on a change of coordinates
which are non-sinusoidal,\footnote{We note here that the Coulomb
potential can in fact be treated using sinusoidal coordinate,
which we shall present in the Appendix.} which according to the
discussions in \cite{Ho} can only be quasi-exactly solvable. This
is consistent with the Lie-algebraic approach to quasi-exactly
models, as these four models, unlike the other six systems, cannot
be obtained as the exactly solvable limits of some quasi-exactly
solvable systems \cite{Tur,GKO}.

The potentials of these later four systems are usually given as
follows \cite{Cooper}:
\begin{eqnarray}
\begin{array}{ll}
\mbox{Coulomb} &:
A\left(A-1\right)\frac{1}{x^2}-2B\frac{1}{x};\\
\mbox{Eckart} &: A\left(A-\alpha\right){{\rm cosech}^2~\alpha x}
-2B{\rm coth}~\alpha x; \\
\mbox{Rosen-Morse II} & : -A\left(A+\alpha\right){{\rm sech}^2~\alpha x}
+2B{\rm tanh}~\alpha x; \\
\mbox{Rosen-Morse I} &:  A\left(A-\alpha\right){{\rm
cosec}^2~\alpha x}+2B{\rm cot}~\alpha x,
\end{array}\label{4pot}
\end{eqnarray}
where $A,~B$ and $\alpha$ are real constants.  Despite the
similarity of the functional forms of the potentials, their
eigenvalues and eigenfunctions, these models are normally being
treated separately in the literature.

The purpose of this paper is to show how these four systems can be
generated by a simple extension of the prepotential presented in
\cite{Ho}.  What is more, they can be treated in a unified manner.
Put simply, the required extension is simply to allow the
coefficients in the prepotential, which are assumed to be
constants in \cite{Ho}, to be dependent on the number of nodes of
the wave functions in such a way that the coefficients of all
terms involving $z$ in the potential are real constants.

To bring out the close similarity of these four potentials, we
find it convenient to express the functions ${\rm cosech},~{\rm
sech}$ and $\rm cosec$ in terms of ${\rm coth},~{\rm tanh}$ and
$\rm cot$, respectively, in (\ref{4pot}).  The resulted forms of
the potentials, the relevant change of coordinates, and their
eigenvalues are listed in Table~1.  For clarity of presentation,
we adopt the unit system in which $\hbar$ and the mass $m$ of the
particle are such that $\hbar=2m=1$.  Also, without loss of
generality, we have absorbed the scale factor $\alpha$ into $x$,
or equivalently, we set $\alpha=1$.

\begin{table}[!]
\caption{\label{table1} The four exactly solvable models based on
non-sinusoidal coordinates. The potential $V(x)$, its relevant
non-sinusoidal coordinate $z(x)$, the derivative $z^\prime(x)$,
and the eigenvalues $E_N$ ($N=0,1,\ldots$) are listed.  Without
loss of generality, we absorb $\alpha$ into $x$, or equivalently,
we set $\alpha=1$. The range of $x$ is: $x\in [0,\infty)$ for the
Coulomb and Eckart potentials, $x\in (-\infty,\infty)$ for the
Rosen-Morse II potential, and $x\in [0,\pi]$ for the Rosen-Morse I
potential.}
\begin{ruledtabular}
\begin{tabular}{ccccc}
 & $V(x)$ & $z(x)$ & $z^{\prime}(x)$ & $E_N$\\
 \hline\\
Coulomb & $A\left(A-1\right)\frac{1}{x^2}-2B\frac{1}{x}$&
$\frac{1}{x}$
 & $-z^2$ & $-\frac{B^2}{\left(A+N\right)^2}$\\
\\
\hline\\
Eckart & $A\left(A-1\right){{\rm coth}^2 x}-2B{\rm coth}~x$ &
${\rm coth}~x$
 & $1-z^2$ & $-\frac{B^2}{\left(A+N\right)^2}-A\left(2N+1\right)-N^2$\\
\\
\hline\\
Rosen-Morse II & $A\left(A+1\right){{\rm tanh}^2 x}+2B{\rm
tanh}~x$ & ${\rm tanh}~x$
 & $1-z^2$ & $-\frac{B^2}{\left(A-N\right)^2}+A\left(2N+1\right)-N^2$\\
\\
\hline\\
Rosen-Morse I & $A\left(A-1\right){{\rm cot}^2 x}+2B{\rm cot}~x$ &
${\rm cot}~x$
 & $-1-z^2$ & $-\frac{B^2}{\left(A+N\right)^2}+A\left(2N+1\right)+N^2$\\
\\
\end{tabular}
\end{ruledtabular}
\end{table}

From the table, we see that these four models involve a change of
variable $z(x)$ whose derivative is of the form
\begin{eqnarray}
z^\prime &=&\lambda - z^2, \label{z}\\
\lambda &=& \left\{
\begin{array}{ll}
               0, &\mbox{: Coulomb;}\\
                1,  & \mbox{: Eckart,~Rosen-Morse II;}\\
               -1, & \mbox{: Rosen-Morse I,}
               \end{array}
               \right.
\end{eqnarray}
and that the potentials can be cast in the same form
\begin{eqnarray}
V(x)=A(A-1)z^2(x) -2B z(x).\label{V-z}
\end{eqnarray}
 In this form, the
four potentials are regular in the variable $z$, with singularity
only at $z=\infty$ and/or $z=-\infty$. As the function $z^{\prime
2}$ is a forth-degree polynomial in $z$, the coordinate $z(x)$ is
non-sinusoidal.  Naively any system based on such coordinate will
be quasi-exactly solvable according to the discussion in
\cite{Ho}.   In the rest of the paper, we shall show how the
exactly solvable models in (\ref{V-z}) can be obtained  from
(\ref{z}).

The organization of the paper is as follows.  In Sect.~II we
review the main points of the prepotential approach of \cite{Ho}
relevant to our present discussions. Sect.~III shows how the
exactly solvable models (\ref{V-z}) can be derived from the
non-sinusoidal coordinates (\ref{z}) by a simple extension of the
approach in \cite{Ho}. The four specific models are then discussed
in detail in Sect.~IV to VII. Sect.~VIII concludes the paper.  In
the Appendix, we show how the Coulomb system is treated based on
sinusoidal coordinate.

\section{Prepotential approach}

The essence of the prepotential approach is as follows. Consider a
wave function $\phi_N(x)$ ($N$: non-negative integer) which is
defined as
\begin{eqnarray}
\phi_N(x)\equiv e^{-W_0(x)}p_N (z),\label{phi_N}
\end{eqnarray}
with
\begin{eqnarray}
p_N (z)\equiv \left\{
               \begin{array}{ll}
                1, & N=0;\\
                \prod_{k=1}^N (z-z_k),& N>0.
                        \end{array}
               \right.
\label{phi}
\end{eqnarray}
Here $z=z(x)$ is some real function of the basic variable $x$,
$W_0(x)$ is a regular function of $z(x)$, and $z_k$'s are the
roots of $p_N(z)$. The variable $x$ is defined on the full line,
half-line, or finite interval, as dictated by the choice of
$z(x)$, We have assumed that the only singularities of the system
are $z=\infty$ and/or $z=-\infty$, as is the case for the four
models of concerned here. The function $p_N(z)$ is a polynomial in
an $(N+1)$-dimensional Hilbert space with the basis $\langle
1,z,z^2,\ldots,z^N \rangle$. $W_0(x)$ defines the ground state
wave function.

 We rewrite $\phi_N$ as
\begin{eqnarray}
 \phi_N =\exp\left(- W_N(x,\{z_k\})
\right), \label{f2}
\end{eqnarray}
with $W_N$ given by
\begin{eqnarray}
W_N(x,\{z_k\}) = W_0(x) - \sum_{k=1}^N \ln |z(x)-z_k|. \label{W}
\end{eqnarray}
Operating on $\phi_N$ by the operator $-d^2/dx^2$ results in a
Schr\"odinger equation $H_N\phi_N=0$, where
\begin{eqnarray}
H_N &=&-\frac{d^2}{dx^2} + V_N,\\
V_N&\equiv&  W_N^{\prime 2} - W_N^{\prime\prime}.
\end{eqnarray}
Hence the potential $V_N$ is defined by $W_N$, and we shall call
$W_N$ the $N$th order prepotential.  From Eq.~(\ref{W}), one finds
that $V_N$ has the form $V_N=V_0+\Delta V_N$:
\begin{eqnarray}
V_0 &=&W_0^{\prime 2} - W_0^{\prime\prime},\nonumber\\
 \Delta V_N &=&
-2\left(W_0^\prime z^\prime
-\frac{z^{\prime\prime}}{2}\right)\sum_{k=1}^N \frac{1}{z-z_k} +
\sum_{k,l\atop k\neq l} \frac{z^{\prime 2}}{(z-z_k)(z-z_l)}.
\label{V}
\end{eqnarray}

Thus the form of $V_N$, and consequently its solvability, are
determined by the choice of $W_0(x)$ and $z^{\prime 2}$ (or
equivalently by $z^{\prime\prime}=(dz^{\prime 2}/dz)/2$). Let
$W_0^\prime z^\prime$ and $z^{\prime 2}$ be taken as polynomials
in $z$. In \cite{Ho}, it was shown that if the degree of
$W_0^\prime z^\prime$ is no higher than one, and the degree of
$z^{\prime 2}$ no higher than two, then in $V_N(x)$ the parameter
$N$ and the roots $z_k$'s, which satisfy the so-called Bethe
ansatz equations (BAE) to make the potential analytic, will only
appear in an additive constant and not in any term involving
powers of $z$. Such system is then exactly solvable. If the degree
of one of the two polynomials exceeds the corresponding upper
limit, the resulted system is quasi-exactly solvable.

Now we are interested in constructing a system based on a
transformed variable $z(x)$ which is a solution of $z^\prime
=\lambda - z^2$.  Since in this case the degree of $z^{\prime 2}$
is four, a direct application of the arguments in \cite{Ho} would
mean that no exactly-solvable system can be generated with
whatever choice of $W_0$.

But it turns out that with a slight extension of the methods in
\cite{Ho}, one can generate the four potentials in the
prepotential approach.  The main observation is this. The
classification of solvability given in \cite{Ho} is valid as long
as the coefficients of the powers of $z$ in $W_0^\prime z^\prime$
are constants which are independent of $N$. If one allows these
coefficients to be $N$-dependent, then it may be possible that in
the $V_N$ obtained, all the coefficients of terms involving powers
of $z$ are $N$-independent constants, thus giving rise to an
exactly-solvable system.  This is in fact the case for the four
systems mentioned above.  We shall demonstrate this in what
follows.

\section{Exactly-solvable models with non-sinusoidal coordinates}

Let us choose $z$ from a solution of $z^\prime=\lambda-z^2$, and
take
\begin{eqnarray}
W_0^\prime =A_1 z+ A_0, \label{W0}
\end{eqnarray}
where $A_1$ and $A_0$ are real parameters.  With this choice of
$W_0^\prime$ and $z^{\prime 2}$, we obtain from (\ref{V})
\begin{eqnarray}
V_0=A_1(A_1+1)z^2 + 2A_1A_0 z + A_0^2 -\lambda A_1
\end{eqnarray}
and
\begin{eqnarray}
\Delta V_N = -\left(\lambda-z^2\right) \left\{
2\left[(A_1+1)z+A_0\right]\sum_{k=1}^N\frac{1}{z-z_k} -
\left(\lambda-z^2\right)\sum_{k,l\atop k\neq l}
\frac{1}{(z-z_k)(z-z_l)}\right\}.
\end{eqnarray}
Using the identities
\begin{eqnarray}
\sum_{k=1}^N \frac{z}{z-z_k} &=&  \sum_{k=1}^N
\frac{z_k}{z-z_k} + N,\\
\sum_{k,l=1\atop k\neq l}^N \frac{1}{(z-z_k)(z-z_l)} &=& 2
\sum_{k,l=1\atop k\neq l}^N
\frac{1}{z-z_k}\left(\frac{1}{z_k-z_l}\right),\\
\sum_{k,l=1\atop k\neq l}^N \frac{z^2}{(z-z_k)(z-z_l)}
  &=& 2 \sum_{k,l=1\atop k\neq l}^N
\frac{1}{z-z_k}\left(\frac{z_k^2}{z_k-z_l}\right)+ N(N-1),
\end{eqnarray}
we rewrite $\Delta V_N$ as
\begin{eqnarray}
\Delta V_N = -\left(\lambda-z^2\right)
\left\{N\left(2A_1+N+1\right)
+2\sum_{k=1}^N\frac{1}{z-z_k}\left[\left(A_1+1\right)z_k+A_0 +
\sum_{l\neq k} \frac{z_k^2-\lambda}{z_k-z_l}\right]\right\}
\end{eqnarray}
To remove the poles in $\Delta V_N$, we must demand that $z_k$'s
satisfy the BAE
\begin{eqnarray}
\sum_{l\neq
k}\frac{z_k^2-\lambda}{z_k-z_l}+\left(A_1+1\right)z_k+A_0
 =0, ~~k=1,2,\dots,N. \label{BAE1}
\end{eqnarray}
With these $z_k$'s, only the first term in $\Delta V_N$ remains,
and the potential $V_N$ becomes
\begin{eqnarray}
V_N=\left(A_1 +N\right)\left(A_1 +N +1\right)z^2 + 2A_1A_0 z +
A_0^2 -\lambda \left[\left(2N+1\right)A_1 +
N\left(N+1\right)\right].\label{V_N}
\end{eqnarray}
Now it is seen that $N$ appears in the coefficient of $z^2$ term,
and thus $V_N$ represents a quasi-exactly solvable system, if
$A_0$ and $A_1$ are some fixed constants.

But in this case it is easy to obtain an exactly solvable
potential. Let us choose $A_1$ and $A_0$ such that the
combinations $A\equiv -(A_1+N)$ and $B\equiv -A_1A_0$ are
$N$-independent real constants. The choice of the signs in the
definitions of $A$ and $B$ are for convenience and are not
essential for the moment (they will have to be determined by the
normalizability of the wave functions for physical systems).
Consequently, $A_1$ and $A_0$ depend on $N$:
\begin{eqnarray}
A_1=-\left(A+N\right),~~A_0=\frac{B}{A+N}.\label{A}
\end{eqnarray}
Then the potential $V_N$ becomes $V_N(x)=V(x)-E_N$, where
\begin{eqnarray}
V(x)=A\left(A-1\right)z^2(x) -2B z(x),\label{V0}
\end{eqnarray}
as advertised in Sect.~I, and
\begin{eqnarray}
 E_N =
 -\frac{B^2}{\left(A+N\right)^2}-\lambda\left[A\left(2N+1\right)
 +N^2\right].\label{E_N}
\end{eqnarray}
Now $V(x)$ is independent of $N$, and can be taken to be the
potential of an exactly-solvable system, with eigenvalues $E_N$
($N=0,1,2,\ldots$).  The corresponding wave functions $\phi_N$ are
given by (\ref{phi_N}) together with (\ref{W0}) and (\ref{A}):
\begin{eqnarray}
\phi_N\sim e^{\left(A+N\right)\int^x dx
z(x)-\frac{B}{A+N}x}~p_N(x),~~N=0,1,\ldots \label{phi_N2}
\end{eqnarray}
From (\ref{BAE1}) and (\ref{A}), the BAE satisfied by the roots
$z_k$'s are
\begin{eqnarray}
\sum_{l\neq k}\frac{z_k^2-\lambda}{z_k-z_l} -\left(A+N-1\right)z_k
+\frac{B}{A+N} =0, ~~k=1,2,\dots,N. \label{BAE2}
\end{eqnarray}

In the sections below, we will show how the four potentials
mentioned emerge from (\ref{V0}) with different $\lambda$.  But
first a comment on the choice of the form of $V(x)$ is in order.
One notes that there is an ambiguity in the definitions of $V(x)$
and $E_N$ in $V_N=V-E_N$:  $V_N$ is invariant under $V\to
V-\alpha$ and $E_N\to E_N-\alpha$ for any real $\alpha$.  This
amounts to the choice of the zero point  of $V(x)$.   The form
$V^{\rm SUSY}$ adopted in supersymmetric quantum mechanics (e.g.,
in \cite{Cooper}), corresponds to the choice $\alpha=E_0$ so that
the ground state energy is zero, i.e. $E_0=0$.  In our case,
$V^{\rm SUSY}$ is obtained from the zero-th order prepotential
$W_0(N=0)$ with $N=0$ (remember now that $A_0$ and $A_1$ in $W_0$
depend on $N$):
\begin{eqnarray}
&&W_0^\prime(N=0) =-Az+\frac{B}{A},\nonumber\\
 &&V^{\rm SUSY}=W_0^{\prime 2}(N=0) -
W_0^{\prime\prime}(N=0)\\
&&~~~~~~~~~= A(A-1)z^2 -2B z + \frac{B^2}{A^2}+ \lambda
A.\nonumber
\end{eqnarray}
The energies are
\begin{equation}
E^{\rm SUSY}_N=\lambda\left[A^2
-\left(A+N\right)^2\right]+\frac{B^2}{A^2}-\frac{B^2}{\left(A+N\right)^2}.
\end{equation}
In the literature on supersymmetric quantum mechanics,
$W_0^\prime(N=0)$ is generally called the superpotential.

\section{$\boldmath{\lambda=0}$: the Coulomb case}

In this case, $\lambda=0$ and $z^\prime =-z^2$.  A solution is
$z=1/x$.  The domain of $x$ is $x\in[0,\infty)$. The potential is
\begin{equation}
V(x)=\frac{A(A-1)}{x^2} -2\frac{B}{x}.\label{V-C}
\end{equation}
From (\ref{phi_N2}) the wave function is
\begin{eqnarray}
\phi_N(x)\sim x^{A+N}e^{-\frac{B}{A+N}x}~p_N(z).\label{wf-C}
\end{eqnarray}
For $\phi_N\to 0$ at $x=0$ and $\infty$, one must have $A>0$ and
$B>0$.  The eigen-energies are given by (\ref{E_N})
\begin{eqnarray}
E_N = -\frac{B^2}{\left(A+N\right)^2}, ~~N=0,1,\ldots\label{E-C}
\end{eqnarray}
The roots $z_k$'s satisfy the BAE (\ref{BAE2}) with $\lambda=0$.
It should be emphasized that the above expressions are valid for
any real $A>0$ (not necessary integer) and $B>0$.

The ordinary Coulomb potential is obtained by setting $A=l+1$
($l=0,1,\ldots$) and $B=e^2/2$, where $l$ is the orbital angular
quantum number and $e$ the electric charge. This gives the well
known form of the Coulomb energies $E_N=-e^2/4(N+l+1)^2$. To write
the wave functions (\ref{wf-C}) in the familiar form, we express
$z$ and $z_k$ in $p_N(z)$ in terms of $x=1/z$ and $x_k=1/z_k$.
This gives (by factoring out all $z$'s and $z_k$'s)
\begin{eqnarray}
\phi_N(x)\sim x^{l+1}e^{-\frac{e^2}{2(N+l+1)}x}\prod_{k=1}^N
\left(x-x_k\right), ~~N=1,2,\ldots
\end{eqnarray}
Recall that $p_N(x)=1$ for $N=0$. The BAE satisfied by $x_k$'s are
obtained from (\ref{BAE2}) by setting $\lambda=0$ and changing
$z_k, z_l$ to $1/x_k, 1/x_l$. The final result can be written as
\begin{eqnarray}
\sum_{l\neq k}\frac{1}{x_k-x_l} +\frac{l+1}{x_k}
=\frac{e^2}{2(N+l+1)}, ~~k=1,2,\dots,N.\label{BAE-C}
\end{eqnarray}
Letting $y\equiv e^2x/(N+l+1)$, we can simplify the BAE to
\begin{eqnarray}
\sum_{l\neq k}\frac{1}{y_k-y_l} +\frac{\gamma/2}{y_k}=\frac{1}{2},
~~k=1,2,\dots,N,\label{BAE-Lag}
\end{eqnarray}
where $\gamma\equiv 2(l+1)$.  Eq.~(\ref{BAE-Lag}) is just the set
of equations satisfied by the roots of the Laguerre polynomials
$L_N^{\gamma-1} (y)$ \cite{Szego,CS}.  Hence in terms of the
variable $y$, the wave functions are
\begin{eqnarray}
\phi_N (y)\sim
y^{l+1}e^{-\frac{y}{2}}L_N^{2l+1}\left(y\right),\label{wf-C1}
\end{eqnarray}
which are exactly the form given in Table 4.1 of \cite{Cooper}.

\section{$\boldmath{\lambda=1}$: the Eckart case}

We take $z={\rm coth}~x$.  The range of $x$ is again the half-line
$x\in[0,\infty)$.  The potential, energies and wave functions are
\begin{eqnarray}
V(x)&=&A\left(A-1\right){\rm coth}^2 x -2 B {\rm coth}~x,\nonumber\\
E_N&=&-\frac{B^2}{\left(A+N\right)^2} -A\left(2N+1\right)-N^2,\\
\phi_N&\sim& \left({\rm
sinh}~x\right)^{A+N}e^{-\frac{B}{A+N}x}~p_N\left(z\right).\nonumber
\end{eqnarray}
Now the boundary conditions  $\phi_N\to 0$ as $x\to 0$ and
$x\to\infty$ require $A>0$ and $B>(A+N)^2$, respectively. Hence
$B$ must be greater than $A^2$, i.e. $B>A^2$, in order to admit at
least one bound state (corresponding to $N=0$).  For fixed $A$ and
$B>A^2$, the maximal value of $N$ is such that $B>(A+N)^2$ remains
valid. Thus the number of bound states is $N+1$.

Since $z_k^2\neq 1$ one can divide (\ref{BAE2}) (with $\lambda=1$)
by $z_k^2-1$. Writing the result in partial fractions, we obtain
\begin{eqnarray}
\sum_{l\neq k}\frac{1}{z_k-z_l}
+\frac{\frac{1}{2}\left(\alpha+1\right)}{z_k-1}
+\frac{\frac{1}{2}\left(\beta+1\right)}{z_k+1} =0,
 ~~k=1,2,\dots,N,\label{BAE-Jac}
\end{eqnarray}
with
\begin{equation}
\alpha=-A-N+\frac{B}{A+N},~~\beta=-A-N-\frac{B}{A+N}.
\end{equation}
 One
recognizes that (\ref{BAE-Jac}) are the equations satisfied by the
roots of the Jacobi polynomial $P^{(\alpha,\beta)}_N(z)$
\cite{Szego,CS}. Hence the wave functions for the Eckart potential
are
\begin{eqnarray}
\phi_N\sim \left({\rm
sinh}~x\right)^{A+N}e^{-\frac{B}{A+N}x}~P^{(\alpha,\beta)}_N(z).
\end{eqnarray}
It is easy to check that $\phi_N$ can be expressed as
\begin{eqnarray}
\phi_N(x)\sim \left(z-1\right)^{\frac{\alpha}{2}}
\left(z+1\right)^{\frac{\beta}{2}} P^{(\alpha,\beta)}_N(z).
\end{eqnarray}
This is the form presented in \cite{Cooper}.

\section{$\boldmath{\lambda=1}$: the Rosen-Morse II case}

Here $\lambda=1$ is the same as for the Eckart potential, but now
we take a different solution $z(x)={\rm tanh}~x$.  This choice
implies the variable $x$ is defined on the whole line, $-\infty
<x<\infty$. The wave functions are
\begin{eqnarray}
\phi_N\sim \left({\rm
cosh}~x\right)^{A+N}e^{-\frac{B}{A+N}x}~p_N\left(z\right).
\end{eqnarray}
The boundary conditions  $\phi_N\to 0$ as $|x|\to \infty$ lead to
\begin{eqnarray}
A+N<0,~~|B|<\left(A+N\right)^2.\label{AB}
\end{eqnarray}
So in this case $A$ must be negative, and $B$ can have any sign as
long as the inequality in (\ref{AB}) for $B$ is satisfied.  For
easy comparison with the corresponding expressions in
\cite{Cooper}, we change $A\to -A$ and $B\to -B$. With these new
definitions of $A$ and $B$, the potential (\ref{V0}) and energies
(\ref{E_N}) are written as
\begin{eqnarray}
V(x)&=&A\left(A+1\right){\rm tanh}^2 x + 2 B {\rm tanh}~x\nonumber\\
E_N&=&-\frac{B^2}{\left(A-N\right)^2} + A\left(2N+1\right)-N^2,\\
&& A>0,~~|B|<\left(A-N\right)^2.\nonumber
\end{eqnarray}
For fixed $A$ and $B$ the maximal value of $N$ is such that
$|B|<\left(A-N\right)^2$ remains valid.  The roots $z_k$'s now
satisfy a BAE which is the same as that in the Eckart case, but
with the changes $A\to -A$ and $B\to -B$. Hence the wave functions
are
\begin{eqnarray}
\phi_N\sim \left({\rm
cosh}~x\right)^{-(A-N)}e^{-\frac{B}{A-N}x}~P^{(\alpha,\beta)}_N(z),
\end{eqnarray}
where
\begin{equation}
\alpha=A-N+\frac{B}{A-N},~~\beta=A-N-\frac{B}{A-N}.
\end{equation}
Again, one can rewrite $\phi_N$ in the form given in \cite{Cooper}
\begin{eqnarray}
\phi_N(x)\sim \left(1-z\right)^{\frac{\alpha}{2}}
\left(1+z\right)^{\frac{\beta}{2}}~ P^{(\alpha,\beta)}_N(z).
\end{eqnarray}

We note there that in this case the two models defined by only a
difference in the sign of $B$ are simply mirror images of each
other: they are related by the parity transformation $x\to -x$.

\section{$\boldmath{\lambda=-1}$: the Rosen-Morse I case}

Finally we come to the Rosen-Morse potential defined with the
solution $z=\cot{x}$ for $\lambda=-1$.  The system is defined on
the finite interval $x\in [0,\pi]$.  From the wave function
\begin{eqnarray}
\phi_N\sim \left(
\sin{x}\right)^{A+N}e^{-\frac{B}{A+N}x}~p_N\left(z\right),
\end{eqnarray}
one infers that $A>0$ while $B$ is arbitrary. Again in order to
facilitate comparison with the expressions in \cite{Cooper}, we
change $B\to -B$.  The two systems differing only in the signs of
$B$ are equivalent by reflection and periodicity.

The potential and energies read
\begin{eqnarray}
V(x)&=&A\left(A-1\right)\cot^2 x + 2 B \cot{x}\nonumber\\
E_N&=&-\frac{B^2}{\left(A+N\right)^2}+ A\left(2N+1\right)+N^2.
\end{eqnarray}
The BAE (\ref{BAE2}) is
\begin{eqnarray}
\sum_{l\neq k}\frac{z_k^2+1}{z_k-z_l} -\left(A+N-1\right)z_k
-\frac{B}{A+N} =0,  ~~k=1,2,\dots,N. \label{BAE-RMI}
\end{eqnarray}
This set of BAE is related to the Jacobi polynomials, but with
imaginary argument.  To see this, let us rewrite (\ref{BAE-RMI})
in terms of $y\equiv iz$.  One gets
\begin{eqnarray}
\sum_{l\neq k}\frac{1}{y_k-y_l}
+\frac{\frac{1}{2}\left(\alpha+1\right)}{y_k-1}
+\frac{\frac{1}{2}\left(\beta+1\right)}{y_k+1} =0,
 ~~k=1,2,\dots,N,\label{BAE-RMI-y}
\end{eqnarray}
where
\begin{equation}
\alpha=-A-N-i\frac{B}{A+N},~~\beta=-A-N+i\frac{B}{A+N}.
\end{equation}
Hence the wave functions are
\begin{eqnarray}
\phi_N\sim \left(\sin{x}\right)^{A+N}e^{\frac{B}{A+N}x}~
P^{(\alpha,\beta)}_N\left(iz\right).
\end{eqnarray}
This is consistent with the expression in \cite{Cooper} in the
form
\begin{eqnarray}
\phi_N\sim
\left(z^2+1\right)^{-\frac{A+N}{2}}e^{\frac{B}{A+N}\cot^{-1}z}~
P^{(\alpha,\beta)}_N\left(iz\right).
\end{eqnarray}

\section{Summary}

In \cite{Ho} a new approach to both exact and quasi-exact
solvabilities was proposed.  In this approach the solvability of a
one-dimensional quantum system can be solely classified by two
integers, the degrees of two polynomials which determine the
change of variable and the zero-th order prepotential.  It was
shown that exactly solvable models can only involve a change of
variable which is sinusoidal, otherwise the system is
quasi-exactly solvable.  As such, four out of the ten exactly
solvable models classified by shape invariance in supersymmetric
quantum mechanics \cite{Cooper}, namely, the Coulomb, Eckart,
Rosen-Morse type I and II models, are not covered by the approach
as presented in \cite{Ho}.  In this paper, we have shown how these
four models could be easily generated in a unified way in the
prepotential approach by allowing the coefficients in the
prepotential, which are assumed to be constants in \cite{Ho}, to
be dependent on the number of nodes of the wave functions.

Thus with the results presented in this paper and those in
\cite{Ho}, all the well known one-dimensional exactly solvable
Schr\"odinger models have been generated by the prepotential
approach in a way which, we believe, is much simpler and direct
than other approaches.

\begin{acknowledgments}

This work is supported in part by the National Science Council
(NSC) of the Republic of China under Grant Nos. NSC
96-2112-M-032-007-MY3 and NSC 95-2911-M-032-001-MY2. Part of the
work was done during my visit to the Yukawa Institute for
Theoretical Physics (YITP) at the Kyoto University supported under
NSC Grant No. 97-2918-I-032-002.  I would like to thank R. Sasaki
and the staff and members of YITP for their hospitality.  I am
also grateful to Y. Hosotani for useful discussion and
hospitality.

\end{acknowledgments}

\appendix*

\section{Coulomb potential in sinusoidal coordinates}
Of the four potentials discussed in the main text, Coulomb
potential can be treated in terms of sinusoidal coordinate,
namely, $z(x)=x$ (as in the main text, for simplicity, we absorb
any scale factor into $x$).  Unlike those discussed in the main
text, in the present case what we want is to generate a potential
that is singular at $x=0$. According to \cite{Ho}, the
prepotential $W_0(x)$ should be modified to ${\tilde
W}_0(x)=W_0(x)-A\ln|x|$, where $A$ is some real parameter, to
ensure proper behavior of the wave function at the singular point
$x=0$. Naively, from the discussions n \cite{Ho}, the degree of
$W_0^\prime z^\prime$ should not be more than one in order to get
exactly solvable model.  So let us try $W_0^\prime(x)=b$
($z^\prime=1$). Thus the ground state wave function is $\phi_0\sim
\exp(-{\tilde W}_0)\sim x^A\exp(-bx)$. Normalizability of $\phi_0$
requires $A>0$ and $b>0$.

Replacing $W_0$ in (\ref{V}) by ${\tilde W}_0$, we obtain
\begin{eqnarray}
V_0=\frac{A(A-1)}{x^2} - \frac{2Ab}{x} + b^2
\end{eqnarray}
and
\begin{eqnarray}
\Delta V_N = -\frac{2A}{x} \sum_{k=1}^N \frac{1}{x_k}
+2\sum_{k=1}^N\frac{1}{x-x_k}\left[\sum_{l\neq k}
\frac{1}{x_k-x_l}+\frac{A}{x_k}-b\right].
\end{eqnarray}
Simple poles are removed if $x_k$'s satisfied the BAE
\begin{equation}
\sum_{l\neq k}
\frac{1}{x_k-x_l}+\frac{A}{x_k}-b=0,~~k=1,2\ldots,N.\label{BAE-C1}
\end{equation}
Summing over $k$ in (\ref{BAE-C1}) gives
\begin{equation}
A\sum_{k=1}^N\frac{1}{x_k}-bN=0,
\end{equation}
and hence $\Delta V_N =-2bN/x$.  Finally, we have
\begin{eqnarray}
V_N=\frac{A(A-1)}{x^2} - 2\frac{b(A+N)}{x} + b^2.
\end{eqnarray}
Since $N$ appears in the $1/x$ term in $V_N$, the system is
quasi-exactly solvable in this form.  But as discussed in the main
text, one can make this system exactly solvable by allowing $b$ to
depend on $N$ in such a way that the coefficient of $1/x$ term is
$N$-independent, i.e.
\begin{equation}
b=\frac{B}{A+N},
\end{equation}
where $B$ is a real constant. So the Coulomb potential is given by
\begin{eqnarray}
V_N=\frac{A(A-1)}{x^2} - 2\frac{B}{x} +
\frac{B^2}{\left(A+N\right)^2}.
\end{eqnarray}
This is consistent with the results given in (\ref{V-C}) and
(\ref{E-C}). The BAE (\ref{BAE-C1}) becomes
\begin{equation}
\sum_{l\neq k}
\frac{1}{x_k-x_l}+\frac{A}{x_k}=\frac{B}{A+N},~~k=1,2\ldots,N.
\end{equation}
Letting $A=l+1$ and $B=e^2/2$, we recover the BAE in
(\ref{BAE-C}), and subsequently the wave function $\phi_N$ in
(\ref{wf-C1}).

\end{document}